\documentclass[journal=nalefd,manuscript=letter]{achemso}

\usepackage{chemformula} 
\usepackage[T1]{fontenc} 

\author{Nuttawut Kongsuwan}
\altaffiliation{N.K.and X.X. contributed equally to this work.}
\affiliation[Imperial]{The Blackett Laboratory, Prince Consort Road, Imperial College London, London SW7 2AZ, United Kingdom}

\author{Xiao Xiong}
\altaffiliation{N.K.and X.X. contributed equally to this work.}
\author{Ping Bai}
\author{Jia-Bin You}
\author{Ching Eng Png}
\author{Lin Wu}
\email{wul@ihpc.a-star.edu.sg}
\affiliation[Astar]{Institute of High Performance Computing, A*STAR (Agency for Science, Technology and Research), 1 Fusionopolis Way, \#16-16 Connexis, Singapore 138632, Singapore}

\author{Ortwin Hess}
\email{o.hess@imperial.ac.uk}
\affiliation[Imperial]{The Blackett Laboratory, Prince Consort Road, Imperial College London, London SW7 2AZ, United Kingdom}

\title[Quantum Plasmonic Immunoassay]{Quantum Plasmonic Immunoassay Sensing}

\abbreviations{AuNP, NPoM, QNM, FoM, PL, cQED}

\begin{document}




\begin{abstract}
Plasmon-polaritons are among the most promising candidates for next generation optical sensors due to their ability to support extremely confined electromagnetic fields and empower strong coupling of light and matter. Here we propose quantum plasmonic immunoassay sensing as an innovative scheme, which embeds immunoassay sensing with recently demonstrated room temperature strong coupling in nanoplasmonic cavities. In our protocol, the antibody-antigen-antibody complex is chemically linked with a quantum emitter label. Placing the quantum-emitter enhanced antibody-antigen-antibody complexes inside or close to a nanoplasmonic (hemisphere dimer) cavity facilitates strong coupling between the plasmon-polaritons and the emitter label resulting in signature Rabi splitting. Through rigorous statistical analysis of multiple analytes randomly distributed on the substrate in extensive realistic computational experiments, we demonstrate a drastic enhancement of the sensitivity up to nearly 1500\% compared to conventional shifting-type plasmonic sensors. Most importantly and in stark contrast to classical sensing, we achieve in the strong-coupling (quantum) sensing regime an enhanced sensitivity that is no longer dependent on the concentration of antibody-antigen-antibody complexes -- down to the single-analyte limit. The quantum plasmonic immunoassay scheme thus not only leads to the development of plasmonic bio-sensing for single molecules but also opens up new pathways towards room-temperature quantum sensing enabled by biomolecular inspired protocols linked with quantum nanoplasmonics.
\end{abstract}

\section{Introduction}
Plasmon-polaritons facilitate strong light-matter interaction by squeezing light into subwavelength volumes \cite{maier2007plasmonics,Novotny:2012}. 
It triggers a wide range of nanoscale phenomena and applications such as integrated nano-circuits \cite{xiong2013silver,bermudez2015coupling}, nanolasers \cite{bergman2003surface,hill2014advances,Pickering:2014:10.1038/ncomms5972}, ultraslow (broadband) waves \cite{Tsakmakidis:2017:10.1126/science.aan5196}, surface-enhanced Raman spectroscopy \cite{ding2016nanostructure}, and plasmonic sensing \cite{arroyo2016non}.
Many efforts have been focused on the Purcell enhancement of spontaneous emission rate \cite{xd_1946purcell}, however, remaining in the weak coupling regime. In this weak-coupling regime, the energy decays fast and before a significant coherent interchange between plasmon-polaritons and emitters takes place. Recent experiments and computational simulations have successfully demonstrated \cite{vakevainen2013plasmonic,antosiewicz2014plasmon,zengin2015realizing,chikkaraddy2016single,liu2017strong,Grob:2018} and explained \cite{Kongsuwan:2018} room-temperature strong light-matter coupling, offering the possibilities for quantum information processing at room temperature and in ambient environment \cite{altewischer2002plasmon,gonzalez2011entanglement,tame2013quantum,xu2018quantum}. 
Indeed, the field of nanoplasmonic strong coupling is rapidly evolving from emitter-ensemble \cite{vakevainen2013plasmonic,antosiewicz2014plasmon,zengin2015realizing} toward single-emitter strong coupling \cite{chikkaraddy2016single,liu2017strong,Grob:2018} which is of key importance in many quantum technologies \cite{birnbaum2005photon,kasprzak2010up,vasa2017strong}, heralding various functional devices operating at the single photon level \cite{chang2007single,hensen2017strong,tiecke2014nanophotonic,Sun57}. Here, we link ensemble-, few- and single-emitter nanoplasmonic strong coupling with immunoassay sensing. 

An immunoassay is a biochemical test that measures the presence or concentration of specific molecules in a solution using antibodies \cite{wild2005immunoassay}. A molecule detected in an immunoassay is often referred to as an ``analyte''. Clearly, the ultimate goal of immunoassay sensing is to detect a single analyte enabling, for example, diagnosis of early-stage cancer. 
However, due to the size mismatch between typical analytes (typically $<$100 nm) and the optical wavelength (400$-$700 nm), single-analyte sensing has remained elusive. Hence, most immunoassays rely on specific detectable labels that are chemically linked with antibodies. In a plasmonic immunoassay, the sensing signal can be further enhanced as plasmonic structures can efficiently squeeze light into tiny, sub-wavelength volumes that are comparable to the size of the analyte.
Various types of labels that are chemically linked to antibodies, can further enhance the sensing signals via their interactions with plasmon-polaritons.
For example, a dielectric label has been used to induce changes in the refractive index, shifting the plasmonic resonance \cite{wang2011magnetic,krishnan2011attomolar}. 
In addition to the refractive index change, a metallic nanoparticle label can also induce hybridization of plasmonic modes with the surrounding structure \cite{lyon1998colloidal,hong2012contribution,szunerits2014surface,yang2016exploiting}. Fluorophores have also been proposed as sensing labels in surface-plasmon field-enhanced fluorescence spectroscopy \cite{liebermann2000surface,wang2009prostate,wang2016directional} where the photoluminescence is enhanced due to the Purcell effect of the plasmonic resonance. In spite of this significant enhancement through a suitable nanoplasmonic environment, the sensing process -- interactions between various types of labels and plasmon-polaritons in modern plasmonic immunoassays -- have to date not yet explored the strong-coupling quantum regime. 

In this study, we propose a new scheme that embeds and utilizes strong coupling between quantum emitter label(s) and plasmon-polaritons to achieve a drastically enhanced sensitivity, down to a single-analyte quantum plasmonic immunoassay. It is from this single-analyte perspective that we shall in the following start our discussion before subsequently moving to the experimentally more common multi-analyte (ensemble) case. 
In each case, the ultra-sensitivity is achieved via the characteristic spectral signature of  Rabi splitting which is effectively a bi-directional shift. Compared to label-free plasmonic sensing, the presence of quantum emitter labels provides a sensitivity enhancement of a factor of 14.2, while solitary dielectric and gold nanoparticle (AuNP) labels give sensitivity enhancements of only 2.62 and 2.73, respectively. 

Our statistical studies on multi-analyte detection demonstrate that the proposed protocol also works well in the weak coupling regime (just as conventional plasmonic sensors do) when the emitter labels are displaced from the plasmonic hotspot. In the case of multiple randomly positioned analytes, the optical spectrum of the composite system is not necessarily associated with a Lorentzian line-shape and we consequently introduce a figure of merit (FoM) as the integral of the spectral changes. The immunoassay-FoM in the weak coupling (or classical) regime decreases from 0.226 to 0.093 with decreasing number of analytes. In contrast, the immunoassay-FoM in the strong coupling (or quantum) regime remains approximately constant (around 0.360), independent of the concentration of emitter labels. Indeed, it retains this value also for a single emitter label.

\begin{figure}[t]
\centering
\includegraphics[width=1\linewidth]{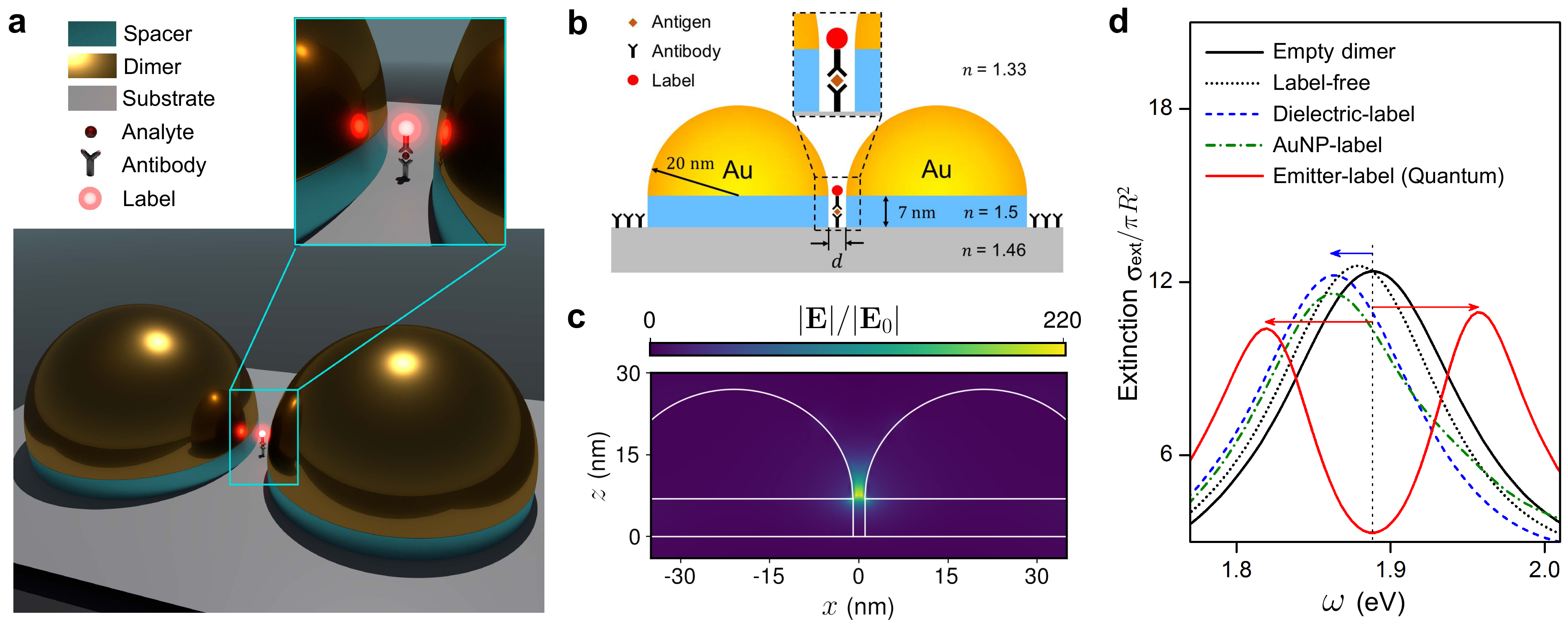}
\caption{Quantum plasmonic immunoassay sensing. (a) Schematic illustration of the strong-coupling immunoassay setup. A gold hemisphere nano-dimer cavity captures an immunoassay complex in the proximity of the plasmonic hotspot. (b) Side-view of a representative system with a 20nm-radius hemisphere placed on top of the dielectric spacer with a thickness of 7 nm, which matches the thickness of antibody-antigen-antibody complex. (c) Illustration of normalized electric field hotspot located between the  distribution inside the dimer. (d) Performance comparisons of different plasmonic sensors. The extinction cross-sections are normalized by the geometric cross-section of the hemisphere $\pi R^2$.}
\label{fig1}
\end{figure}

\section*{Results and discussion}
\subsection*{Strong-Coupling Immunoassay}

The principle and set-up of our proposed quantum plasmonic immunoassay sensing protocol are schematically illustrated in Fig. \ref{fig1}a. It functionally involves four main parts: 
(i) a plasmonic nano-dimer cavity (here formed by two gold hemispheres),
(ii) an antigen as the ``analyte'' to be detected, 
(iii) a sensing label (which is used to enhance the sensitivity in either a classical or quantum regime), and 
(iv) two antibodies which are paired with the target antigen and chemically linked with the sensing label, respectively. 
For our strong-coupling immunoassay, we use a quantum emitter (such as a quantum dot) as the sensing label, and the term ``strong-coupling" refers to the strong coupling between the emitter label and the plasmonic field in the cavity. 
The sensing label can, however, also be a dielectric or plasmonic nanoparticle resulting in a conventional (classical) plasmonic immunoassay. 
We generally assume that immunoassay takes place in a liquid environment, such as water or serum. 
A critical layer of dielectric spacer is introduced between the plasmonic hemispheres and the substrate to provide the opportunity to adjust the vertical position of the hotspot, $e.g.$, to coincide with the sensing label. Figure \ref{fig1}b shows the side-view for a representative strong-coupling immunoassay system.  
The dimer gap between the two nanoplasmonic hemispheres is denoted as $d$. All the simulations are performed on the basis of a 3D full-wave spatio-temporal simulation method based on a finite-difference time-domain (FDTD) scheme  (Lumerical Solutions Inc.; see Methods).

To appreciate the functionality of the various components, let us first establish the optical response of an empty hemisphere dimer. Figure \ref{fig1}c shows the normalized electric field distribution of a dimer with a gap $d$ = 2 nm resulting in a plasmonic resonance at 1.89 eV. The characteristic plasmonic hotspot is clearly seen inside the nanogap, with a field enhancement by a factor of 195 at the gap center. It is this enhancement that generally forms the foundation for plasmonic sensing with high sensitivity in an ambient environment.
Now to demonstrate the sensitivity of the strong-coupling immunoassay, Fig. \ref{fig1}d compares its extinction spectrum (solid red curve) to the spectra of label-free (dotted black curve), dielectric-labeled  (dashed blue curve) and AuNP-labeled (dash-dotted green curve) immunoassays. In all cases of the study, we use the same gold hemisphere dimer and dielectric spacers but vary the different sensing complexes inside the gap. 

In the label-free case,  we place only one antibody and one antigen inside the gap, adjusting the height of the antibody such that the antigen is close to the plasmonic hotspot, while in the case of the labeled complexes we adjust the geometry such that the label resides close to the plasmonic hotspot. The dielectric label itself is approximated as a simple nano-cylinder (radius: 1 nm and height: 2 nm) with a refractive index of 1.8, whereas the AuNP label is modeled as a gold nano-sphere (with the diameter of 1 nm). We use a full-wave spatio-temporal Maxwell-Bloch model  \cite{Kongsuwan:2018} to take into account the energy exchange dynamics between the two-level quantum emitters (linewidth of 26 meV) and the plasmonic field, revealing the dynamics in both weak- and strong-coupling regimes. Compared to conventional sensing protocols that have a characteristic shift of the optical resonance, the strong-coupling immunoassay exhibits the characteristic signature of strong coupling, $i.e.$, the two Rabi peaks, as indicated in Fig. \ref{fig1}d. This is in effect a bi-directional shift with higher sensitivity than those of conventional immunoassays. In the following, we will investigate and present the principles of the strong-coupling immunoassay.

\subsection*{Anti-Crossing of Strong Coupling}
\begin{figure}[!ht]
\centering
\includegraphics[width=0.8\linewidth]{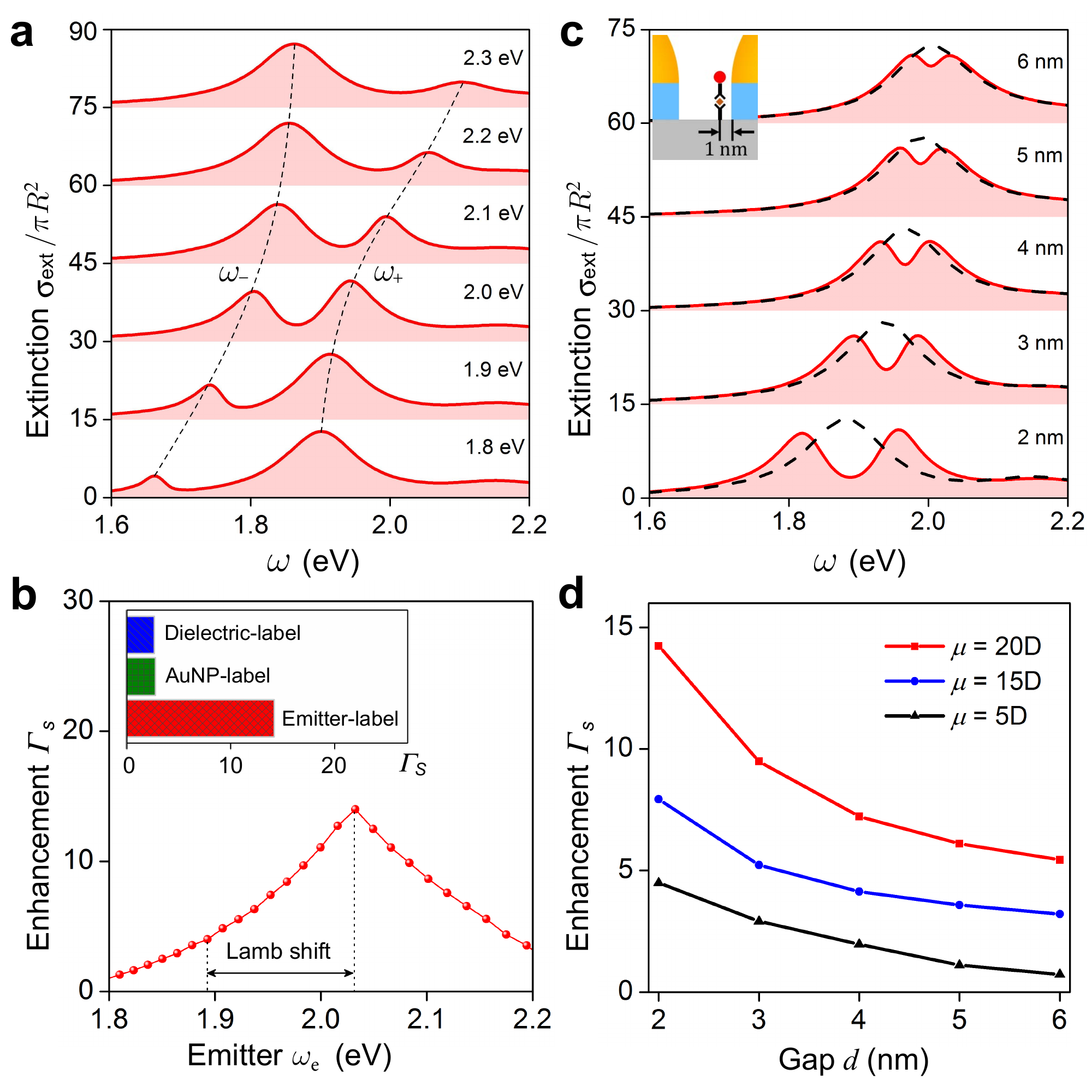}
\caption{Strong-coupling anti-crossing of emitter label(s) and plasmon-polaritons. (a) Spectrum of extinction cross-sections $\sigma_\mathrm{ext}$ for emitter resonance of $\omega_{\mathrm{e}}$ = 1.8--2.3 eV, with plasmonic resonance fixed at $\omega_{\mathrm{p}}$ = 1.89 eV. Two dashed curves denoted as branches $\omega_+$ and $\omega_-$ are plotted for the guidance of Rabi splitting.
(b) Sensitivity enhancement $\Gamma_{\mathrm{S}}$ as a function of the emitter resonance $\omega_{\mathrm{e}}$, with transition dipole moment $\mu$ = 20D. Inset: comparison of sensitivity enhancement $\Gamma_{\mathrm{S}}$ for plasmonic sensors with different types of labels. (c) Spectrum of extinction cross-sections $\sigma_\mathrm{ext}$ for dimers with $d$ = 2--6 nm, with transition dipole moment $\mu$ = 20D. As the dimer gap $d$ increases, the plasmonic resonance (black dashed lines) shifts, so the emitter resonance is also tuned accordingly. Inset: zoom-in view of the gap region, with the distance between emitter label and the hemisphere on the right fixed at 1 nm. (d) Sensitivity enhancement $\Gamma_{\mathrm{S}}$ as a function of the gap size $d$, with transition dipole moment $\mu$ = 5D, 10D, 20D, respectively.
The spectra in (a) and (c) are shifted upward in steps of 15 for clarity.}
\label{fig2}
\end{figure}

For an emitter coupled with a plasmonic resonance, the system-Hamiltonian can be written as \cite{baranov2017novel}:
\begin{eqnarray}
\mathcal{H}=
\begin{pmatrix}
\omega_{\mathrm{e}}-i\gamma&g\\
g&\omega_{\mathrm{p}}-i\kappa
\end{pmatrix},
\end{eqnarray}
where $\omega_{\mathrm{e}}$ and $\omega_{\mathrm{p}}$ are the resonances of the emitter and the plasmon-polaritons, and $\gamma$ and $\kappa$ are their decay rates, respectively. Then the coupled system has two new eigen-frequencies expressed as \cite{baranov2017novel}:
\begin{eqnarray}
\omega_{\pm}(\Delta)=\frac{\omega_{\mathrm{e}}+\omega_{\mathrm{p}}}{2}\pm \mathrm{Re}\left[\sqrt{\frac{1}{4}(\Delta-i(\gamma-\kappa))^2+g^2}\right],
\label{eq2}
\end{eqnarray}
where $\Delta=\omega_{\mathrm{e}}-\omega_{\mathrm{p}}$ is the detuning which can be adjusted to control the optical responses $\omega_{\pm}$ of the coupled system. When the coupling rate is slower than the decay processes $g<|\gamma-\kappa|/2$, the coupled system is in the weak coupling regime where the emitter's emission rate is enhanced by plasmon-polaritons due to the Purcell effect. The system reaches the strong-coupling regime when the coupling exceeds the decay $g>|\gamma-\kappa|/2$, at which the emitter and the plasmon-polaritons become hybridized and coherently exchange energy. 
In this regime, the hybridized system frequencies $\omega_+$ and $\omega_-$ reveal a clear anti-crossing pattern as we sweep the emitter resonance $\omega_{\mathrm{e}}$ across the plasmonic resonance $\omega_{\mathrm{p}}$. As shown in Fig. \ref{fig2}a, the numerically calculated extinction spectra of coupled system show clearly two branches, when the hemisphere dimer with $\omega_{\mathrm{p}}$ = 1.89 eV is coupled to different emitter with $\omega_{\mathrm{e}}$ = 1.8-2.3 eV.
We shall in the following use this type of spectral signature to characterize the performance of our strong-coupling immunoassay scheme.

\subsection*{Sensitivity}
For conventional shifting-type plasmonic sensors ($e.g.$, the dotted black, dashed blue, and dash-dotted green curves in Fig. \ref{fig1}d), the (sensing) sensitivity is generally defined as the ratio of the change in sensor output ($e.g.$, resonance shift $\delta\omega$) relative to the change in the quantity to be measured ($e.g.$, concentration of the analyte $\delta c$) \cite{homola2008surface}. However, while $\delta c$ is a good measure in the case of appreciable analyte concentrations, measuring concentrations is clearly no longer well-applicable in the context of few- or single-analyte detection. We thus adopt the number of analytes $N$ (where $N=$ 1, 2, 3, ...) as quantifying descriptor to define sensitivity as: 
\begin{eqnarray}
\textsc{S}^{\mathrm{shift}}_{N}=\frac{\delta\omega}{N},
\end{eqnarray} to be used here to characterize the shifting-type plasmonic sensors for few- or single-analyte detection, an example of which is indicated by the blue arrow in Fig. \ref{fig1}d.
On the other hand, for the strong-coupling immunoassay (splitting-type), there are double shifts toward both directions, as illustrated by the red arrows in Fig. \ref{fig1}d. 
The measurable frequency splitting between the peaks $\delta \omega = |\omega_+ -\omega_-|$ can be (analytically) expressed as:
\begin{eqnarray}
\delta \omega=\mathrm{Re}[\sqrt{(\Delta-i(\gamma-\kappa))^2+4g^2}].
\label{eq3}
\end{eqnarray}
Compared to the original resonance detuning $|\Delta|$, the system response is modified by $\delta \omega-|\Delta|$ due to the coupling between the emitter(s) and the plasmon-polaritons. Therefore, we define the sensitivity of our strong-coupling (or splitting-type) immunoassay as:
\begin{eqnarray}
\textsc{S}^{\mathrm{split}}_N=\frac{\delta \omega-|\Delta|}{N}.
\end{eqnarray}
We note that in principle, when the emitter and plasmons are on resonance ($\Delta =0$), the frequency difference $\delta \omega = \sqrt{4g^2-(\gamma-\kappa)^2}$ corresponds to the Rabi splitting, and the sensitivity $\textsc{S}^{\mathrm{split}}_N$ should thus reach its maximum. For the off-resonance cases ($|\Delta|>0$), we expect a drop of the sensitivity $\textsc{S}^{\mathrm{split}}_N$.
In the context of single-analyte detection ($N=1$), the sensitivity is purely relevant to the resonance shift for shifting-type sensors ($\textsc{S}^{\mathrm{shift}}_{N=1}=\delta\omega$) or to the resonance splitting for strong-coupling sensors ($\textsc{S}^{\mathrm{split}}_{N=1} = \delta \omega-|\Delta|$).

To directly compare various different types of sensors, we normalize the sensitivities to that of the label-free sensor, denoting the normalized sensitivity as $\Gamma_{\mathrm{S}}=\textsc{S}^{\mathrm{label}}_{N}/\textsc{S}^{\mathrm{label-free}}_{N}$, which characterizes the sensitivity enhancement induced by the sensing labels. The inset of Fig. \ref{fig2}b shows the sensitivity enhancement (extracted from the spectra in Fig. \ref{fig1}d) for different labels. We observe a clear enhancement of 15-fold ($\Gamma_{\mathrm{S}}=14.2$) in strong-coupling immunoassay sensing, making it a competitive candidate for next-generation sensors. We also extract the sensitivity enhancement for the strong-coupling immunoassay from Fig. \ref{fig2}a, which is dependent on the emitter resonance $\omega_{\mathrm{e}}$ as shown in Fig. \ref{fig2}b. 
Clearly, there exists an optimal $\Gamma_{\mathrm{S}}$  as we sweep the emitter resonance. However, due to the optical Lamb shift experienced by the emitter label (see Section S1 in Supporting Information), this optimal $\Gamma_{\mathrm{S}}$ does not occur at $\omega_{\mathrm{p}}$ = 1.89 eV when the emitter and the plasmon-polaritons are on resonance. This suggests that the ideal emitter label should be slightly detuned from the plasmonic cavity resonance in order to maximize the sensitivity. In practice, however, it might be more convenient to design a suitably detuned plasmonic cavity for any pre-selected emitter label in a plasmonic immunoassay system.   

When seeking to optimize the sensitivity of our immunoassay system we can vary geometrical parameters such as the gap width $d$ or choose a particular emitter label with a characteristic transition dipole moment $\mu$. Determining the dependence of the sensitivity enhancement $\Gamma_{\mathrm{S}}$ on the dimer gap $d$ we first focus on the spectral response of empty hemisphere dimers of variable gap width $d$ = 2--6 nm as shown in Fig. \ref{fig2}c (black dashed lines). As the dimer gap $d$ decreases, the plasmonic resonance is red-shifted, thus the emitter resonance must be tuned for each gap $d$ according to the similar optimization procedure illustrated in Fig. \ref{fig2}b. The corresponding extinction spectra with optimized $\omega_{\mathrm{e}}$ and maximized $\Gamma_{\mathrm{S}}$ are obtained for various dimer gaps as shown in Fig. \ref{fig2}c (red solid lines). For smaller dimer gap $d$, the Rabi splitting of the coupled system becomes more prominent, indicating a stronger coupling strength between the emitter and the plasmon-polaritons. Similarly, we extract the sensitivity enhancement $\Gamma_{\mathrm{S}}$ from these extinction spectra. As shown in Fig. \ref{fig2}d, $\Gamma_{\mathrm{S}}$ increases drastically as the dimer gap $d$ decreases. Note that, for this gap size study, the emitter is displaced from the gap center and stays 1 nm away from one of the hemisphere's surface, as illustrated in the inset of Fig. \ref{fig2}c. This position is optimized to provide the highest sensitivity enhancement (see Section S2 in Supporting Information). If we now vary the transition dipole moment $\mu$, Fig. \ref{fig2}d shows that as expected, $\Gamma_{\mathrm{S}}$ increases for larger $\mu$. Typical emitters that are used for strong coupling include methylene blue ($\omega_{\mathrm{e}} \sim$ 1.878 eV, $\mu \sim 4$D) \cite{chikkaraddy2016single,patil2000self}, 
J-aggregates ($\omega_{\mathrm{e}} \sim$ 2.145 eV, $\mu \sim 30$D) \cite{zengin2015realizing,liu2017strong}, and rhodamine ($\omega_{\mathrm{e}} \sim$ 1.741 eV, $\mu \sim 5$D) \cite{delysse1998picosecond,wuestner2010overcoming}. 
Moreover, some III-V quantum dots ($e.g.$, InAs and GaAs) \cite{khitrova2006vacuum} and defects in two-dimensional materials ($e.g.$, WS$_\text{2}$ and graphene) \cite{sie2018valley} have also been used, with their $\mu$ ranging from 10D to 100D. In this context, we should point out that the previously mentioned 15-fold sensitivity enhancement is achieved using a conservative value of $\mu = 20$D. Of course, we may anticipate even higher sensitivity enhancement $\Gamma_{\mathrm{S}}$ with a larger dipole moment $\mu$. 
Besides, this study also clearly confirms the preferred choice of dimers with small gaps for improved sensitivity.

\begin{figure}[!ht]
\centering
\includegraphics[width=0.9\linewidth]{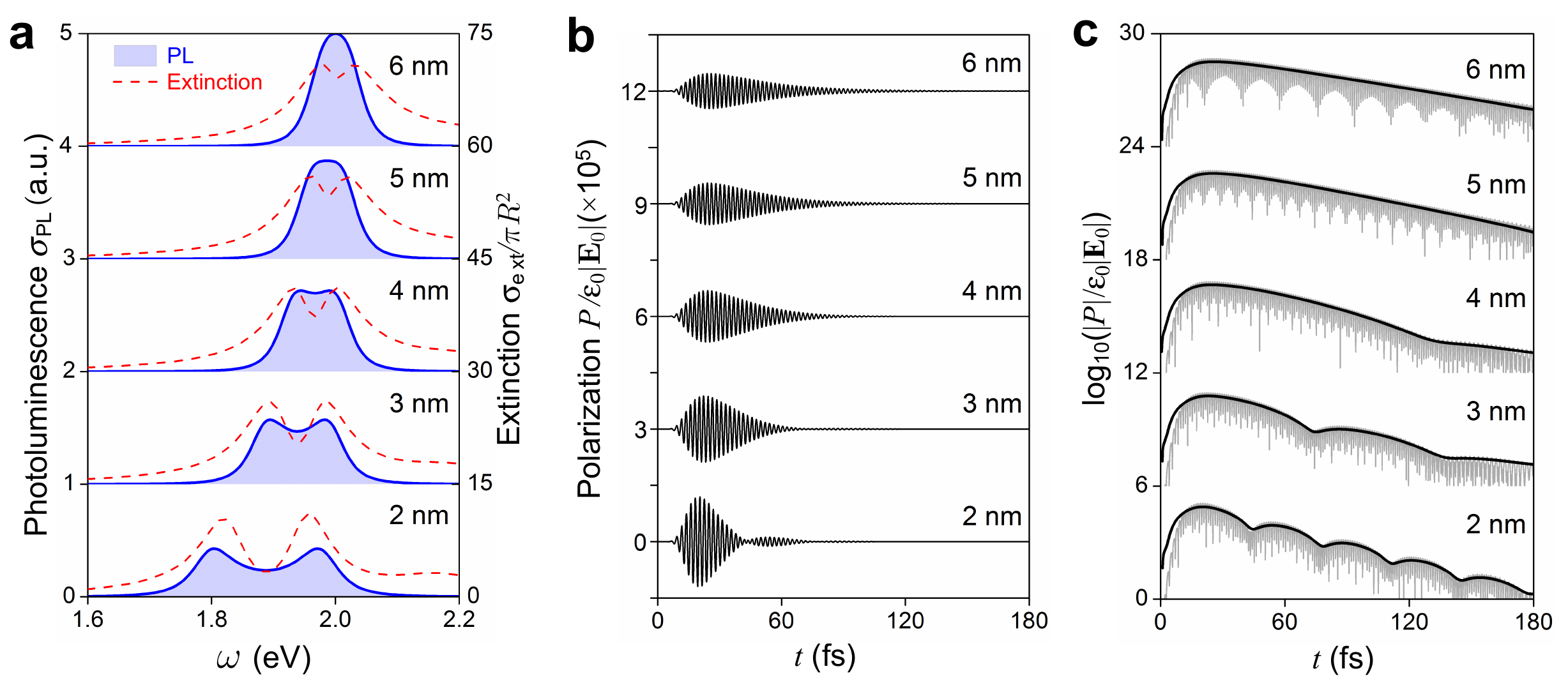}
\caption{Transition from the weak- to strong-coupling regime. (a) Spectrum of photoluminescence $\sigma_{\mathrm{PL}}$ for dimers with $d$ = 2--6 nm and transition dipole moment $\mu$ = 20D. Red dashed lines are the corresponding spectra of extinction cross-sections $\sigma_\mathrm{ext}$. (b) The temporal dynamics of the emitter polarization $P(t)$ for dimers with $d$ = 2--6 nm and transition dipole moment $\mu$ = 20D. (c) The temporal dynamics of polarization $|P(t)|$ (gray curves) and the corresponding envelopes (black curves) in the logarithmic presentation. The curves are shifted upward in steps of (a) 1 for PL and 15 for extinction, (b) 3$\times 10^5$, or (c) 6 for clarity.}
\label{fig3}
\end{figure}

\subsection*{Strong-Coupling Photoluminescence Spectra}
In typical strong coupling experiments, the extinction spectrum is usually obtained by measuring the amount of radiation transmitted through the coupled system as a function of the wavelength of the incident light. However, the ``splitting'' feature manifested in the extinction spectrum is not a sufficient condition to determine the occurrence of strong coupling between one or more emitters and plasmon-polaritons. In fact, such splitting may also originate from absorption or interference \cite{Laussy:2009}.  To rigorously confirm that we are, indeed, achieving strong coupling, we calculate the photoluminescence (PL) spectrum of the emitter in our strong-coupling immunoassay system by deriving a master equation within the framework of cavity quantum electrodynamics (cQED) (see Section S4 in Supporting Information). To represent the Hamiltonian of the coupled system, we adopt the quasi-normal mode (QNM) method and obtain the coupling strength and decay rates from the Purcell factors of QNMs (see Sections S3-S4 in Supporting Information). Experimentally, this PL spectrum can be measured by exciting the emitter with a pump laser and collecting the luminescence from the coupled system \cite{Melnikau:2016,Grob:2018}. Fundamentally, both extinction and PL spectra describe the dynamic interactions of the coupled system. However, they originate from different processes: extinction spectra reflect stimulated absorption and emission processes of the coupled system (whose signal interferes with the background excitation light), whereas the PL reveals the spontaneous emission of the coupled system without the background excitation light.

Figure \ref{fig3}a shows in direct comparison both the PL and extinction spectra for strong-coupling immunoassay systems with $d$ = 2--6 nm. It is for a (large) gap size of $d$ = 6 nm, where the discrepancy between the PL and extinction spectra becomes most noticeable. Although the extinction spectrum has two splitting resonance peaks, the single peak in the PL spectrum indicates that the emitter and plasmon-polaritons are, in fact, in the weak coupling regime. The splitting peaks in the extinction spectrum emerge simply due to the interference between the emitter and plasmon-polaritons \cite{Laussy:2009}.
As the gap size decreases ($d$ $\le$ 4 nm), both the PL and extinction spectra have two split peaks, unambiguously indicating the strong coupling between emitter and plasmon-polaritons. The occurrence of strong coupling is perhaps most strongly manifested in the dynamics of the polarization $P(t)$ of the quantum emitter and its dependence on different gap sizes. Figures \ref{fig3}b and \ref{fig3}c show $P(t)$ normalized by the incident field amplitude $\varepsilon_0|\textbf{E}_0|$ in linear and logarithmic presentations. 
In Fig. \ref{fig3}b, we can see a second wave packet for a small gap $d$ = 2 nm as a direct evidence of the revival of the emitter \cite{rempe1987observation} which occurs thanks to its interaction with the plasmon-polaritons in the strong coupling regime. Such dynamics then emerges in the logarithmic presentation as ripples for small dimer gaps ($d<$ 5 nm).

The PL spectrum and time-dependent analysis reveal more underlying physics than just the extinction spectrum. However, due to complex energy transfer and recombination processes of the vibrational and defect states \cite{Melnikau:2016} involved, it is experimentally also more challenging to observe Rabi splitting in the PL spectrum than in the extinction spectrum. 
It is not less challenging to calculate within a cQED framework the combined PL spectrum of an ensemble of analyte-emitter complexes. As the emitters at different locations experience $e.g.$, different amounts of optical Lamb shifts and Purcell enhancements, the emitters cannot be considered indistinguishable. Indeed, even if they are of the same type the retardations and interactions between different emitters must also be considered. Hence, the computation within the cQED framework would increase exponentially with the emitter number. When in the following analyzing sensing properties of a spatially (randomly) distributed ensemble of analytes we will model the spatio-temporal strong-coupling dynamics in terms of a full-wave Maxwell-Bloch approach \cite{Kongsuwan:2018} adapted to our immunoassay configuration. Adopting an experimentally relevant set-up we will focus on the extinction spectrum which is of course accurate enough in terms of spectroscopically characterizing the sensing performance through comparisons before and after the adsorption of analyte-emitter complexes.

\subsection*{Statistics and Figure of Merit (FoM)}

\begin{figure}[!ht]
\centering
\includegraphics[width=1\linewidth]{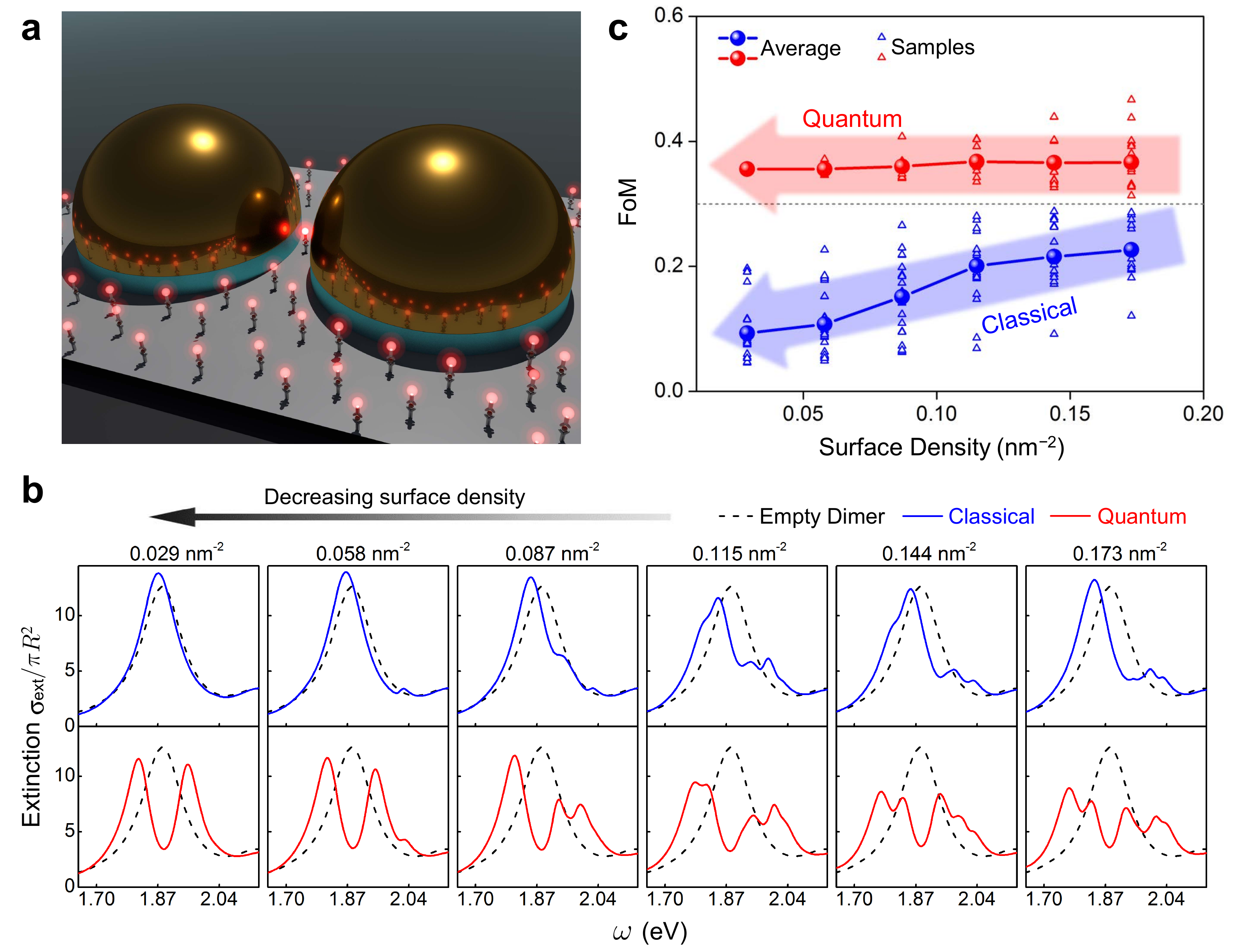}
\caption{Multi-analyte detection with randomly distributed analyte-emitter complexes. (a) Sketch of the multiple analyte-emitter complexes randomly distributed surrounding the hemisphere dimer. The emitter label inside the gap center shall be strongly coupled with plasmonic resonance, while the others remain weakly coupled. (b) Representative spectra of extinction cross-sections $\sigma_\mathrm{ext}$ in classical and quantum regimes, with different surface densities of analyte-emitter complexes. (c) The figure of merit (FoM) as a function of the surface density of analyte-emitter complexes. The hollow triangles represent the FoMs of each simulation sample, and the sphere symbols are the mean values at each surface density. The arrows indicate the change of FoM as the surface density of analyte-emitter complexes decreases.}
\label{fig4}
\end{figure}

In a typical immunoassay set-up, there will be a number of analytes dissolved in a liquid. To realistically model experiments with this set-up in our {\em in silico} experiment (for details regarding the simulation set-up see Section S5 in Supporting Information), we now focus on an ensemble of analyte-complexes (each including a quantum emitter), which are randomly distributed in the vicinity of the nano-dimer and subject to an average surface density. This random distribution reflects the consequence of many complex physical and chemical processes such as transport and molecular inter-interaction which indeed merit consideration in future studies. As illustrated in Fig. \ref{fig4}a, most of the analyte-complexes will in a typical situation be randomly distributed around the two nano-dimer hemispheres, but there is also the chance to find one or more of the complexes in the gap between the hemispheres. Indeed, with increasing surface density of analyte-emitter complexes the probability to find an analyte-emitter in the center of the dimer gap increases. This reflects the fact that just like in an {\em in vitro} experiment, even with all the quantum emitters resonance at $\omega_{\mathrm{e}}$ = 2.03 eV, every emitter experiences a different nano-plasmonic environment and thus exhibits different degrees of the optical Lamb shifts. 

We perform a statistical study on the case of multiple analyte-emitter complexes, and obtain the extinction spectra of 30 samples for each surface density (see Figs. S5-S10 in Supporting Information). From the large simulation sample space, we select several representative spectra and classify them into two groups, as shown in Fig. \ref{fig4}b. For the first group (top row of Fig. \ref{fig4}b), the spectra (solid blue curves) exhibit a major peak shifted from the original plasmonic resonance (dashed black curves), accompanied by some minor peaks, especially when the surface density increases. For these spectra with one major peak, the interaction between emitter label and plasmon-polaritons is still in the weak coupling regime, $i.e.$, none of the emitters is in the center of the dimer gap. In contrast, for the second group (bottom row of Fig. \ref{fig4}b), the spectra (solid red curves) have two major peaks, each of which may possess their own splitting sub-peaks. This ``splitting'' indicates that the strong coupling occurs because of a ``random'' emitter located in the center of the dimer gap. Accordingly, we define the first group (or the shifting-type) and the second group (or the splitting-type) as the classical and quantum sensors, respectively. 

It is worth noting that, as the surface density of analyte-emitter complexes increases, the spectrum in quantum regime no longer consists of well-defined (or Lorentz-shape) peaks at distinct frequencies. Therefore, we define a figure of merit (FoM) to characterize the performance of multi-analyte detection as:
\begin{equation}
\label{eq:fom}
\text{FoM}=\frac{\int |\sigma_\mathrm{ext} - \sigma^0_\mathrm{ext}| \mathrm{d}\omega}{\int \sigma^0_\mathrm{ext} \mathrm{d}\omega},
\end{equation}
in which we integrate the change of extinction cross-section over the whole spectral range and normalize it to the extinction cross-section $\sigma^0_\mathrm{ext}$ of the empty dimer. 
This FoM effectively quantifies the optical change before and after the adsorption of the analyte-emitter complexes.
The statistics on FoM for different surface densities of analyte-emitter complexes are summarized in Fig. \ref{fig4}c, where the hollow triangles represent the samples that we have studied (30 samples for each surface density).
The solid lines show the average FoMs of classical and quantum sensors at each surface density, and the dotted line at FoM = 0.3 is used to differentiate classical and quantum regimes. The histogram of FoMs for all the studied samples can also be found in Fig. S11 in Supporting Information. 

From Fig. \ref{fig4}c, we observe that for a large surface density of analyte-emitter complexes, there is no clear boundary between the FoMs of classical and quantum regimes. 
As the surface density decreases, FoM in the classical regime drops drastically from 0.226 to 0.093. 
However, the FoM in quantum regime remains almost constant ($\sim$ 0.360),
because it dominantly results from the single emitter located at the plasmonic hotspot. 
This proves that, toward single-analyte detection, the strong-coupling immunoassay protocol unambiguously outperforms the shifting-type sensors. 
We also note an increased fluctuation in FoM for quantum sensors with increasing surface density. A larger number of quantum emitters randomly scattered near the plasmonic hotspot introduces uncertainty in the coupling strength due to complex many-body interactions between the emitters and plasmon-polaritons.

So far, we have presented a simple nanostructure with a few-nm gap to demonstrate splitting-type sensing. Such a small size of this gap is clearly not without practical challenges. For example, fabrication of a nanogap must be sufficiently pristine in order to accommodate an antibody, and the steric effect could be significant in such a small gap and prevent a complex from binding at the narrowest (optimal) location. However, our analysis with many different random distributions shows that an analyte does not necessarily have to be at this most narrow point to achieve quantum plasmonic immunoassay sensing. In fact, strong coupling could be achieved at a larger gap size using a quantum emitter with larger dipole moment. Alternatively, an open cavity using a plasmonic nanocube \cite{liu2017strong} could also provide sufficient field confinement at its corners and is certainly more accessible for biomolecules. Another alternative could be detecting a small antigen without a much larger antibody by using artificial proteins to capture the antigen between a plasmonic dimer \cite{chevrel2015specific,yim2008synthesis,gurunatha2016nanoparticles}.

\subsection*{Conclusion} 
In summary, we have proposed and demonstrated the effectiveness of an innovative immunoassay sensing protocol by employing quantum emitters as sensing labels of anitibody-antigen-antibody complexes which in (hemispherical) nanoplasmonic open cavities facilitate room-temperature strong coupling.
The proposed splitting-type sensing approach demonstrates a nearly 15-fold sensitivity enhancement over conventional shifting-type label-free plasmonic sensors. The underlying mechanism of the gigantic sensitivity enhancement is attributed to the splitting of the original resonance, which is equivalent to a bi-directional resonance shifting. 
Using quantum emitter labels with realistically strong transition dipole moments $\mu$, extensive numerical experiments with varying nanogap widths allow us to suggest that nanogaps below 5 nm would be ideal systems for strong coupling conditions and also suggest slight detuning from the plasmonic resonance.
Moreover, using randomly distributed  multi-analyte systems with variable random analyte distributions as realistic model systems for biological experiments reveals that our proposed immunoassay protocol is able to perform both classical and quantum sensing. 
A new figure of merit (FoM) is defined to quantify the spectral difference before and after the adsorption of the analyte-emitter complexes in this multi-analyte study. We show that with decreasing concentration of analytes, the FoM degrades rapidly for classical sensors, yet quantum sensors are able to maintain a stable, concentration-independent FoM.  This demonstrates a genuine potential for experiments based on our quantum plasmonic immunoassay sensing protocol to finally reach the single-analyte detection limit. 
While the current proposal engages a quantum phenomenon ($i.e.$, strong coupling with a quantum emitter) to detect a classical object ($i.e.$, antigen) the approach can conceivably be extended to detect a quantum object such as an electronic spin in nanodiamonds \cite{maze2008nanoscale,taylor2008high,doherty2013nitrogen}.  
This opens up a new pathway toward plasmonics-enabled room-temperature quantum sensing \cite{fan2015quantum,lee2016quantum,RevModPhys.89.035002}.  


\section*{Methods}

To obtain the extinction cross-sections and polarization responses, we perform finite-difference time-domain simulations using a commercial software from Lumerical version 8.18.1298. The permittivity of gold is modeled from the experimental results from Johnson and Christy \cite{Johnson:1972}. The nanostructure is illuminated by the total-field scattered-field (TFSF) source with background refractive index $n_\mathrm{B}=1.33$. The simulation domain size is 1 $\mu$m $\times$ 1 $\mu$m $\times$ 1 $\mu$m. We utilize the conformal meshing scheme between dielectric interfaces, but not on metal interfaces, with a maximum step size of 30 nm in all directions. While finer meshes of 0.25 nm and 0.5 nm are used on analyte-complexes and the Au hemisphere dimer, respectively. 

The dynamics of the (two-level) quantum emitter label(s) is modelled self-consistently with the full-wave spatio-temporal dynamics of the optical field on the basis of an FDTD scheme. An individual emitter is (for simplicity) assumed as embedded within a single Yee cell which interacts with the electromagnetic fields through its polarization density $\mathbf{P}$ in the Maxwell equation $\varepsilon_0 \varepsilon  \partial \mathbf{E}/\partial t=\nabla \times \mathbf{H} - \partial \mathbf{P}/\partial t$. For an emitter with transition frequency $\omega_{\mathrm{e}}$ and transition dipole moment $\mu=$ 20D (unless otherwise stated) oriented along unit vector $\mathbf{n_e}=\mathbf{e_z}$, its polarization $\mathbf{p}$ is calculated by the Maxwell-Bloch equations \cite{Boyd:2003,Kongsuwan:2018}: 
\begin{equation}
\mathbf{p}=\Delta x^3\mathbf{P}=2\mu \mathrm{Re}[\rho_{12}],
\label{eq:p}
\end{equation}
\begin{equation}
\frac{\partial^2 \mathbf{p}}{\partial t^2} + 2\Gamma\frac{\partial \mathbf{p}}{\partial t} + (\omega_{\mathrm{e}}^2+\Gamma^2)\mathbf{p} =\frac{2\omega_{\mathrm{e}}\mu^2}{\hbar}(\rho_{11}-\rho_{22})[(\mathbf{E}-\mathbf{E}^\mathrm{div})\cdot\mathbf{n_e}]\mathbf{n_e},
\label{eq:maxwell_bloch_p}
\end{equation}
\begin{equation}
\frac{\partial \rho_{22}}{\partial t} = - \frac{\partial \rho_{11}}{\partial t} = -\gamma_0 \rho_{22} + \frac{1}{\hbar\omega_{\mathrm{e}}}\left(\frac{\partial \mathbf{p}}{\partial t}+\Gamma\mathbf{p}\right)\cdot(\mathbf{E}-\mathbf{E}^\mathrm{div}),
\end{equation}
where $\Delta x$ = 0.25 nm is the mesh size at the emitter's position, $2\Gamma$ = 26 meV is the FWHM linewidth, 
$\gamma_0=\omega_{\mathrm{e}}^3 n_\mathrm{B}^3 \mu^2/3\pi\varepsilon_0\hbar c^3$ is the relaxation rate,  
and $\rho_{11}$, $\rho_{22}$, $\rho_{12}=\rho_{21}^*$ are the diagonal and off-diagonal elements of the density matrix. Note that in our simulations, the total electric field $\mathbf{E}$ is corrected by the divergent field \cite{Schelew:2017}:
\begin{equation}
\mathbf{E}^\mathrm{div}=-(1+f(\Delta x))\mathbf{p}/3\varepsilon_0 \varepsilon_\mathrm{B} \Delta x^3,
\end{equation}
\begin{equation}
f(\Delta x)=-(3/4\pi)^{2/3}(1.15\omega_{\mathrm{e}}\Delta x^3\sqrt{\varepsilon_\mathrm{B}}/c)^2,
\end{equation}
where $\varepsilon_\mathrm{B}=n_\mathrm{B}^2=1.77$ is the background permittivity, as the polarization field generated by the emitter is through numerical self-interaction in a Yee-cell grid (unless corrected as done in our simulations) re-inserted into the simulation and thus (erroneously) re-excited. 

\section*{Author Information}
\subsection*{Corresponding Authors}
E-mail: wul@ihpc.a-star.edu.sg
\newline
E-mail: o.hess@imperial.ac.uk

\subsection*{Author Contributions}
N.K.and X.X. contributed equally to this work. N.K. and X.X. performed the simulations and wrote the manuscript. N.K. and J.Y. carried out theoretical derivations. P.B., C.E.P., L.W. and O.H. contributed to data analysis. L.W. and O.H. supervised the entire project. All authors commented on the manuscript.

\subsection*{Note}
The authors declare no conflict of interest.

\begin{acknowledgement}
L.W. gratefully acknowledges the fruitful discussions with Prof. Zhang-Kai Zhou from Sun Yat-sen University, China. C.E.P. and L.W. acknowledge financial support from the National Research Foundation Singapore NRF2017-NRF-NSFC002-015, NRF2016-NRF-ANR002; and A*STAR SERC A1685b0005. O.H. acknowledges support from the Engineering and Physical Sciences Research Council (EPSRC) UK through projects EP/L024926/1 and EP/L027151/1. The authors thank Erik Dujardin for many insightful discussions.

\end{acknowledgement}

\begin{suppinfo}
Evaluation of the optical Lamb shift of an emitter label at different locations. Location of the emitter label in the simulations with different gap size. The theory of Quasinormal modes. Multi-mode Purcell enhancement estimations of plasmonic modes. Statistical study of multi-analyte detection with varying analtye concentrations.
\end{suppinfo}

\providecommand{\latin}[1]{#1}
\makeatletter
\providecommand{\doi}
  {\begingroup\let\do\@makeother\dospecials
  \catcode`\{=1 \catcode`\}=2 \doi@aux}
\providecommand{\doi@aux}[1]{\endgroup\texttt{#1}}
\makeatother
\providecommand*\mcitethebibliography{\thebibliography}
\csname @ifundefined\endcsname{endmcitethebibliography}
  {\let\endmcitethebibliography\endthebibliography}{}

\end{document}